\documentclass[runningheads]{llncs}
\usepackage{tikz}
\usepackage{graphicx}
\usepackage{blindtext}
\usepackage{enumitem}
\usepackage{xcolor}
\usepackage{algorithm2e}
\usepackage{bm}
\usepackage{blindtext, graphicx, caption, mathtools}
\usepackage{wrapfig}
\usepackage{booktabs}
\usepackage{todonotes}

\usetikzlibrary{automata, positioning}
\usetikzlibrary{shapes,snakes}
\usetikzlibrary{matrix,arrows,backgrounds}
\begin{document}
\title{Energy Efficiency in Reinforcement Learning for Wireless Sensor Networks\thanks{This study was supported by EPSRC and SPHERE IRC, grant number EP/K031910/1. 
The authors would like to thank all colleagues from EurValve and SPHERE.
This paper was accepted on 30/07/2018 and presented at the ECML-PKDD Workshop Green Data Mining 2018 on 14/09/2018.}}
%
%
\author{Micha\l{} Koz\l{}owski\inst{1} \and
Ryan McConville\inst{1} \and
Ra\'ul Santos-Rodr\'iguez\inst{2} \and
Robert J. Piechocki\inst{1}}
\authorrunning{M. Koz\l{}owski et al.}
%
\institute{Department of Electrical and Electronic Engineering \and
Department of Engineering Mathematics \\ University of Bristol, UK \\
\email{\{m.kozlowski, 
ryan.mcconville, enrsr, r.j.piechocki\}@bristol.ac.uk}}
\maketitle              
\begin{abstract}

As sensor networks for health monitoring become more prevalent, so will the need to control their usage and consumption of energy. This paper presents a method which leverages the algorithm's performance and energy consumption. By utilising Reinforcement Learning (RL) techniques, we provide an adaptive framework, which continuously performs weak training in an energy-aware system.  We motivate this using a realistic example of residential localisation based on Received Signal Strength (RSS). The method is cheap in terms of work-hours, calibration and energy usage. It achieves this by utilising other sensors available in the environment. These other sensors provide weak labels, which are then used to employ the State-Action-Reward-State-Action (SARSA) algorithm and train the model over time. Our approach is evaluated on a simulated localisation environment and validated on a widely available pervasive health dataset which facilitates realistic residential localisation using RSS. We show that our method is cheaper to implement and requires less effort, whilst at the same time providing a performance enhancement and energy savings over time.

\end{abstract}

\section{Introduction}

Sensor networks for pervasive monitoring of health are increasingly common. They serve to extend the knowledge of patient's recovery progress, by collecting and analysing data long after the patient has left a health care facility and has returned home. However, the environmental impact of these systems often takes a secondary place during the design process, surrendering to performance or practicality of deployment. As these systems become more ubiquitous, the energy costs associated with their operation will become an important deciding factor. 

Aided by RL techniques, we propose a new method, designed to alleviate the need for rigorous training and dependence on energy-consuming sensors. We do this by performing weak training across the entirety of the sensor network's lifespan. Additionally, by utilising more power-hungry sensors sporadically, we can achieve continuous improvement of performance while at the same time reducing the need to use them in the first place.

The method is validated by the use of a dataset which was collected within the SPHERE project \cite{twomey_sphere_2016}. The experiments performed as part of this dataset aim to resemble `natural' residential behaviours as closely as possible, by including numerous participants performing scripted and non-scripted actions in a test bed environment. It includes data from bespoke sensors which are popular within the pervasive health community. These include RGB-D cameras, environmental sensors (ES), passive infrared (PIR) and wearable accelerometer and RSS data. ES often act as Access Points (APs) for the wearable RSS.

In this paper, indoor localisation is considered as a case study. Localisation is an important objective when rolling out a health monitoring sensor network \cite{woznowski_sphere:_2017}. Depending upon the sensors used to infer locations, it can also be a significant consumer of the energy budget in the infrastructure \cite{zafari_survey_2017}. The study will follow state-of-the-art localisation techniques as benchmark for our method, presented in detail in \cite{kozlowski_data_2018}.

We aim to provide a reliable and cheap indoor localisation solution capable of adapting to a persistent environment. This adaptability is required as Received Signal Strength-based localisation is notoriously arduous to deploy and unforgiving in a dynamic environment \cite{li_turf:_2017}. The dynamics in this context can be understood as constantly changing radio frequency (RF) signatures due to human or non-human factors. 

The contributions of this paper include:

\begin{enumerate}
  \item Novel, energy-aware adaptive localisation algorithm: We create a simulated environment, closely resembling a real-life localisation system and exhaustively test our method in various simulated experiments.  
  \item Validation on true pervasive health dataset: We show that the algorithm is easy to generalise to different environments, and can be adapted to various localisation models. We do this using data of differing levels of calibration.
  \item Effects of action selection: We compare the effects of different action selection mechanisms in terms of energy-efficiency, and discuss which method is best suited for this purpose. We perform this test on both the simulated and real-life experiments.
\end{enumerate}

The paper is structured as follows: In Section \ref{related} we provide the current state-of-the-art work being done on this topic. Then, in Section \ref{method} we present our methods and discuss their suitability in terms of this problem. Finally, in Sections \ref{evaluation} and \ref{validation}, we present our results, discuss their significance and provide points for future work in \ref{conclusion}.

\section{Related work}\label{related}

The problem set out in this paper was inspired by the work done in \cite{possas_egocentric_2018}. Here, the authors attempt to classify activities of the user by on-body sensors and video cameras. They also consider the energy consumption of the camera, utilising a Markov Decision Process (MDP) to decide whether to use weak, but efficient accelerometer and gyroscope, or strong but inefficient video cameras. 

Let us consider the above study in terms of indoor localisation. The idea of energy-efficient localisation has been proposed before \cite{abu-mahfouz_alwadha_2017,zheng_energy-efficient_2017}. These methods calculate the energy efficiency directly - either by adapting the transmission power to the environment, or by inherently using low-power devices. In our paper we consider energy-efficiency in terms of number of sensors. We aim to reduce the usage of different sensors, with only a broad idea of their power consumption. This makes our method easy to generalise and adapt to already existing sensor networks.

The available pervasive health monitoring sensors, such as PIR or ES \cite{fafoutis_experiences_2018} differ in their usability and the quality of their readings. They also differ in how much energy it takes to operate them and process their results \cite{elsts2018}. Low-power wearable sensors \cite{fafoutis_designing_2017} are also popular within the community, providing not only the on-board acceleration observations, but also acting as a RF anchor. The fusion of these sensors \cite{kozlowski_data_2018} show that different combinations can provide improvement to localisation and activity recognition \cite{possas_egocentric_2018} performance.

During infrastructure deployments, it is vital to perform a `calibration run' or fingerprinting. This method is widely acknowledged in literature \cite{subedi_practical_2017,yiu_wireless_2017}, and involves an offline recording phase and an online localisation phase. Fingerprints can be recorded whilst performing specific activities \cite{twomey_sphere_2016} or simply `living' in the house \cite{byrne_rssi_2018}. Fingerprinting is known to be time-consuming, and is often performed just once \cite{li_turf:_2017}. The latter means that the model is not updated or retrained, even in the case of dynamic environment, which can lead to sub-par performance. 

It was also found that each user will perform differently during deployment calibrations \cite{mcconville_understanding_2018}. Performance of the user can vary considerably, regardless of level of technical knowledge or skill. The model trained by one user may not work optimally on others. This would necessitate the system remembering, or dynamically assigning appropriate models to their users. Whilst the users could theoretically be recognised directly \cite{mcconville_person_2018}, that would involve substantial processing of the sensor data. 

As shown, there is a clear need for an adaptive method of continuous weak learning. We can alleviate the concerns of energy efficiency by making the system aware of its consumption, even in the broadest of terms. Further, this model could be adapted to more complicated energy studies. We can also remove the need for user-specific training by re-estimating the model at certain intervals. The system would be thus indifferent to specific user training, relying instead on weak re-estimation over time to tailor the model to specific users. 

\section{Method}\label{method}
\subsection{Markov Decision Processes and SARSA}
The reinforcement problem is formulated using an MDP. MDPs are tuples of \{$\mathcal{S}, \mathcal{A}, \mathcal{P}, \mathcal{R}$\}, where $\mathcal{S}$ is the state-space, $\mathcal{A}$ is the action space, $\mathcal{P}$ is the transition kernel, $\mathcal{R}$ the immediate reward function. Additionally, we recognise two parameters, $\gamma$ and $\alpha$ -- the discount factor and learning rate respectively. For any MDP, there exists an optimal policy $\bm{\pi^*}: \mathcal{S} \rightarrow \mathcal{A}$. The desired outcome of the MDP is to estimate this policy. In this paper, we utilise the SARSA algorithm defined as follows \cite{sutton_reinforcement_1998}:


\begin{equation}\label{sarsa}
    Q(s_t, a_t) \leftarrow Q(s_t, a_t) + \alpha[r(s_t, a_t) + \gamma Q(s_{t+1}, a_{t+1}) - Q(s_t, a_t)]
\end{equation}\\
where $Q$ is the state-action value matrix, which is updated at each iteration, $\alpha$ is the learning rate, $\gamma$ is the discount factor, and $r \in \mathcal{R}$ is the immediate reward at state $s_t$ and action $a_t$. We assume that the dynamics $\mathcal{P}$ are equally likely for each state, given each action. 

We will now formalise our problem in terms of the above. The reinforcement state space is given by $\mathcal{S} = \{ S1, S2 \}$. These two states specify whether at time $t$ we use `enhanced' or `low-power' sensing. In this paper, $S1$ will signify the `enhanced' sensing, which provides reliable labels at the cost of high-energy usage. This state also allows for the system to perform the re-estimation of the parameters, using the labels which were recently observed by these sensors. `Low-power' sensors will be denoted by $S2$. Accordingly, each state will be able to perform one of two actions $\mathcal{A} = \{ A1, A2 \}$, which in turn lead the system to their respective states. 

The reward function was designed to be simple and intuitive. It penalises the system if it remains in $S1$ and rewards if in $S2$. More formally:

\[
r(s_t, a_t) = 
\begin{cases}
-1 & \text{if $s_t = S1 \quad$ and $\quad a_t = A1$}\\
+1 & \text{if $s_t = S2 \quad$ and $\quad a_t = A2$}\\
0 & \text{else}\\
\end{cases}
\]
Additionally, at each time step the system is rewarded if the performance error is reduced or remains the same, and penalised if it increases. This forces the system to continuously seek performance improvement, even if in S2. The error in these iterations is only calculated during $S1$ from the currently observed labels - in $S2$ the system retains the value from $t-1$. We denote this boost as $B$ and error as $e$:

\[
B_t = 
\begin{cases}
-1 & \text{if $e{(t)} \geq e{(t-1)}$}\\
+1 & \text{if $e{(t)} < e{(t-1)}$}\\
\end{cases}
\]

\begin{figure}[!h]
\centering
\begin{tikzpicture}[auto,node distance=8mm,>=latex,font=\small,squarednode/.style={rectangle, draw=black, thick, minimum size=5mm,fill={rgb:red,16;green,120;blue,150}, text=white}]
    \tikzstyle{round}=[thick,draw=black,circle,fill={rgb:red,235;green,201;blue,54}, text=white]
    \useasboundingbox (-2,-5) rectangle (6.5,2.5);
    \scope[transform canvas={scale=2}]
    \node[round]             (s1) {$S1$};
    \node[round, below right=5mm and 20mm of s1] (s2) {$S2$};
    \node[squarednode, below=of s1] (a1) {$A1$}; 
    \node[squarednode, above=of s2] (a2) {$A2$}; 
    \node (as) at (-0.9, -0.7) {$-1$};
    \node (bs) at (3.4, -0.3) {$+1$};
    
    \node (ab) at (0.8, -0.7) {$0$};
    \node (bb) at (1.7, -0.3) {$0$};
    
    \node (ac) at (1.3, -2.3)  {$0$};
    \node (bc) at (1.4, 1.3) {$0$};

    \draw[->] (s1.west) .. controls (-0.6, -0.7) .. (a1.west);
    \draw[->] (a1.east) .. controls (0.6, -0.7) .. (s1.east);
    \draw[->] (s2.east) .. controls (3.1, -0.3)  .. (a2.east);
    \draw[->] (a2.west) .. controls (1.9, -0.3) .. (s2.west);
    
    \draw[->] (s2.south) .. controls (1.3, -2.1) .. (a1.south);
    \draw[->] (s1.north) .. controls (1.4, 1.1) .. (a2.north);
    \endscope
    
\end{tikzpicture}
\caption{Diagram of the MDP state space.}
\label{fig:diag_mdp}
\end{figure}
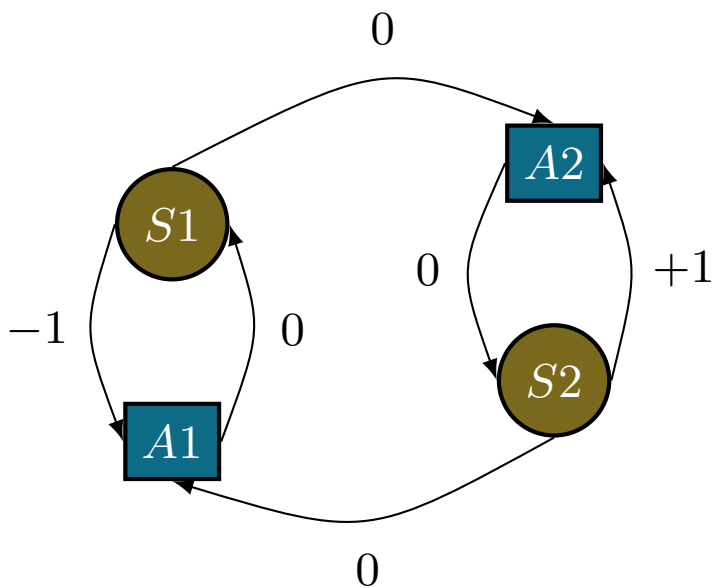

\newpage
\noindent
This reward boost can be trivially added in \eqref{sarsa} as follows:

\begin{equation}
    Q(s_t, a_t) \leftarrow Q(s_t, a_t) + \alpha[B_t + r(s_t, a_t) + \gamma Q(s_{t+1}, a_{t+1}) - Q(s_t, a_t)]
\end{equation}
The MDP environment is shown in Fig. \ref{fig:diag_mdp}. The states are given by circles, the actions are the squares. The numbers next to the arrows specify the reward for each transition. It is crucial to mention that the state space, parametrised by the MDP in Fig. \ref{fig:diag_mdp} differs from the inference state-space, which uses a Hidden Markov Model. The inference space serves to represent physical surroundings, as in \cite{kozlowski_data_2018}, whereas the MDP state-space is an abstract representation of a system state machine.

\subsection{Action Selection}
Selecting the appropriate action for each iteration is not trivial. There exist methods ranging from completely random, pseudo-random and greedy. Greedy selection makes use of the expected future rewards, and exploits them with no regard to any other alternative trajectories, even if the chosen one is sub-optimal. In our approach, the trade-off between exploration and exploitation should be leveraged, such that we converge quickly as to preserve energy, but also retain a degree of exploration, to continue looking for an optimal trajectory. To do this, we use the $\epsilon$-greedy algorithm.

The difference between Greedy and $\epsilon$-Greedy lies in the parameter $\epsilon$. Where Greedy chooses the next action as $Q(s_t, a_t) = max_a Q(s_t, a)$, $\epsilon$-Greedy selects an alternative action with probability $\epsilon$, ensuring that we explore the trajectories more thoroughly in the search of the optimal policy $\bm{\pi^*}$. This is because we no longer `exploit' the reward, prioritising quick convergence, but are open to `explore' the policy space. The larger the parameter $\epsilon$, the broader the exploration, at the cost of higher energy consumption. It is stipulated, that the added adaptability in the form of $\epsilon$, will make $\epsilon$-Greedy better when leveraging efficiency and performance.

 Softmax action selection differs from the above methods, in that sub-optimal choices will be weighted as a graded function of their estimated value. It is likely to reach the optimal policy quicker than Greedy or $\epsilon$-Greedy, but at a cost of higher energy usage. Formally, it chooses action $a$, with probability \cite{sutton_reinforcement_1998}:

\begin{equation}
    P_t = \frac{e^{Q_t(a)/\tau}}{\sum^n_{b=1}e^{Q_t(b)/\tau}}
\end{equation}

In this study we will consider the above three selection methods: Greedy, $\epsilon$-Greedy and Softmax. The usefulness of these methods, given our use case, will be judged by how well they perform in simulation and during validation. 

\subsection{Inference}

The inference method is the same as the one used in previous work \cite{kozlowski_data_2018}. Using standard Bayesian filter notation, we describe the joint probability of the locations $\textbf{x}_t$ and observations $\textbf{z}_t$ as:


\begin{equation} \label{hmm}
p(\boldsymbol{x}_{1:t},\boldsymbol{z}_{1:t}) = p(\boldsymbol{x}_{0})\prod_{i=1}^{t}p(\boldsymbol{z}_{t}|\boldsymbol{x}_{t})p(\boldsymbol{x}_{t}|\boldsymbol{x}_{t-1})
\end{equation}

where $p(\boldsymbol{x}_{0})$ is the prior, $p(\boldsymbol{z}_{t}|\boldsymbol{x}_{t})$ are the emissions and $p(\boldsymbol{x}_{t}|\boldsymbol{x}_{t-1})$ are the transition dynamics of the system. Each state emits observations at time $t$. The symbols in question are sensor observations which we will denote as $z_{t}$. The emitting distributions of these observations is best described by a Gaussian:

\begin{equation} \label{gauss_emit}
p(\boldsymbol{z}_{t}|\boldsymbol{x}_{t}) = \sum_{k=1}^{K}\mathcal{N}(\boldsymbol{z}_{t}|\mu_{jk}, \sigma_{jk})
\end{equation}
where $1 \leq j \leq T $ are the location states, and $1 \leq k \leq M $ is the number of AP sensors.\\

\subsection{Parameter Re-Estimation}

When the system enters $S1$, it is allowed to access to labels from reliable `oracle' sensors. The labels from each `oracle' can be considered as the real prediction of location. Each of these `oracles' maps its output to a given location state. If the `oracles' are activated at time $t$, the system can re-estimate the old emission and transition probabilities with the observations from this `oracle', to which it currently has access. This is done with a single iteration of Expectation-Maximisation (EM) of the previous Gaussian distribution and a sample $o_{t}$. The weighting in EM specifies how much we trust the `oracle' reading -- in essence, it specifies how much of the old distribution should be retained. The optimal weights were found empirically for both the simulation and validation. 

\subsection{Proposed Algorithm}

The following pseudo-code is only intended to serve as an outline of the novel part of the algorithm. With regards to HMM parameter estimation and Forward-Backward inference we refer the reader to the literature, for example \cite{rabiner_fundamentals_1993} and \cite{tang_toward_2010}.

\begin{figure}[ht]
  \centering
    \begin{algorithm}[H]
   \textbf{Input:} $\{\lambda, \mathcal{V}\}$ = HMM parameters, \{$\mathcal{S}, \mathcal{A}, \mathcal{P}, \mathcal{R}, \gamma, \alpha$\} = SARSA parameters\\
      \While{$\mathcal{O}$ available}
      {
        $\mathcal{O} \leftarrow$ vector of sensor observations at t\\
        $\boldsymbol{V} \leftarrow$ infer location $P(\mathcal{V}| \mathcal{O}, \lambda)$\\
        \eIf{$s_t == S1$}
        {
            $\Gamma \leftarrow$ vector of weak oracle labels at t\\
            $\lambda^* \leftarrow$ estimate likelihood $\mathcal{L}(\lambda|\Gamma)$\\ 
            $e(t) \leftarrow$ compare $\boldsymbol{V}$ with $\Gamma$\
        }
        {
            \If{$s_t == S2$}
            {
                $\lambda^* \leftarrow \lambda$\\
                $e(t) \leftarrow e{(t-1)}$\\
            }
        }
        \eIf{$e{(t)} \geq e{(t-1)}$}
        {
            $B_t$ = -1\\
        }
        {
            \If {$e{(t)} < e{(t-1)}$}
            {
                $B_t$ = 1\\
            }
        }
        $a_{t + 1} \leftarrow$ next action based on Greedy, $\epsilon$-Greedy or Softmax and dynamics $\mathcal{P}$\\
        $s_{t + 1} \leftarrow a_{t}$\\
        $Q(s_t, a_t) \leftarrow Q(s_t, a_t) + \alpha[B_t + r(s_t, a_t) + \gamma Q(s_{t+1}, a_{t+1}) - Q(s_t, a_t)]$\\
        $s_{t} \leftarrow s_{t + 1}$\\
        $a_{t} \leftarrow a_{t + 1}$\\
        $\lambda \leftarrow \lambda^*$\\
      }
      \caption{Proposed Algorithm}
      \label{algo}
    \end{algorithm}
  \vspace{-4ex}
\end{figure}

  Algorithm \ref{algo} starts by initialising the HMM and SARSA parameters. Note, that for HMM, $\mathcal{V}$ represents the available state space, whereas $\boldsymbol{V}$ is the inferred, most likely sequence of states. It runs as long as there is data coming from the sensors, shown here as $\mathcal{O}$. We assume that the incoming data stream is vectorised. Each iteration of time $t$ specifies a new vector of incoming data, either collected from `oracle' and RSS sensors, or just RSS. This is dependent upon the state in which the system resided at $t-1$. Inference is performed by running the Forward-Backward algorithm. Depending on the current state of the MDP, the output of this could be compared with the weak labels provided by the `oracles', and the HMM parameters $\lambda$ could be re-estimated. If not, the error is retained from the previous run. The reward boost assignment then follows, again, dependent upon the current state. After choosing next action, with respect to the selection method, SARSA is used to calculate $Q(s_t, a_t)$.

The algorithm will be evaluated on a simulated environment and validated on SPHERE Challenge dataset. In the simulation, we aim to scrutinise the algorithm under comprehensive set of changes in the environment, in order to confirm its capabilities and demonstrate its effectiveness. The validation dataset will serve to verify its usefulness under real-world conditions. 

\section{Evaluation}\label{evaluation}

The simulation setup was created to closely resemble a real-life system. A state-space of varying size was created. Each state $j$ is described in terms of arriving symbols $o_t$ from all $M$ APs. Both the size of the simulation space and the number of simulated APs were incremented. For the simulation space, this changed from 10 to 30, in increments of 10. The size of every state was 1m $\times$ 1m. For APs, the number ranged from 5 to 8. The distributions from each AP were simulated according to a BLE path loss model. The parameters of the model were appropriated from \cite{mellios_off-body_2014}, which was calculated in the same test-bed environment as the SPHERE Challenge dataset. We also define `oracles' in a simulation environment as states, which we observe directly at all time $T$. The amount of `oracle' coverage of the state space was also incremented in 10\% intervals from 10\% to 100\%. 

\begin{figure}[p]
  \centering
  \includegraphics[width=0.6\textwidth]{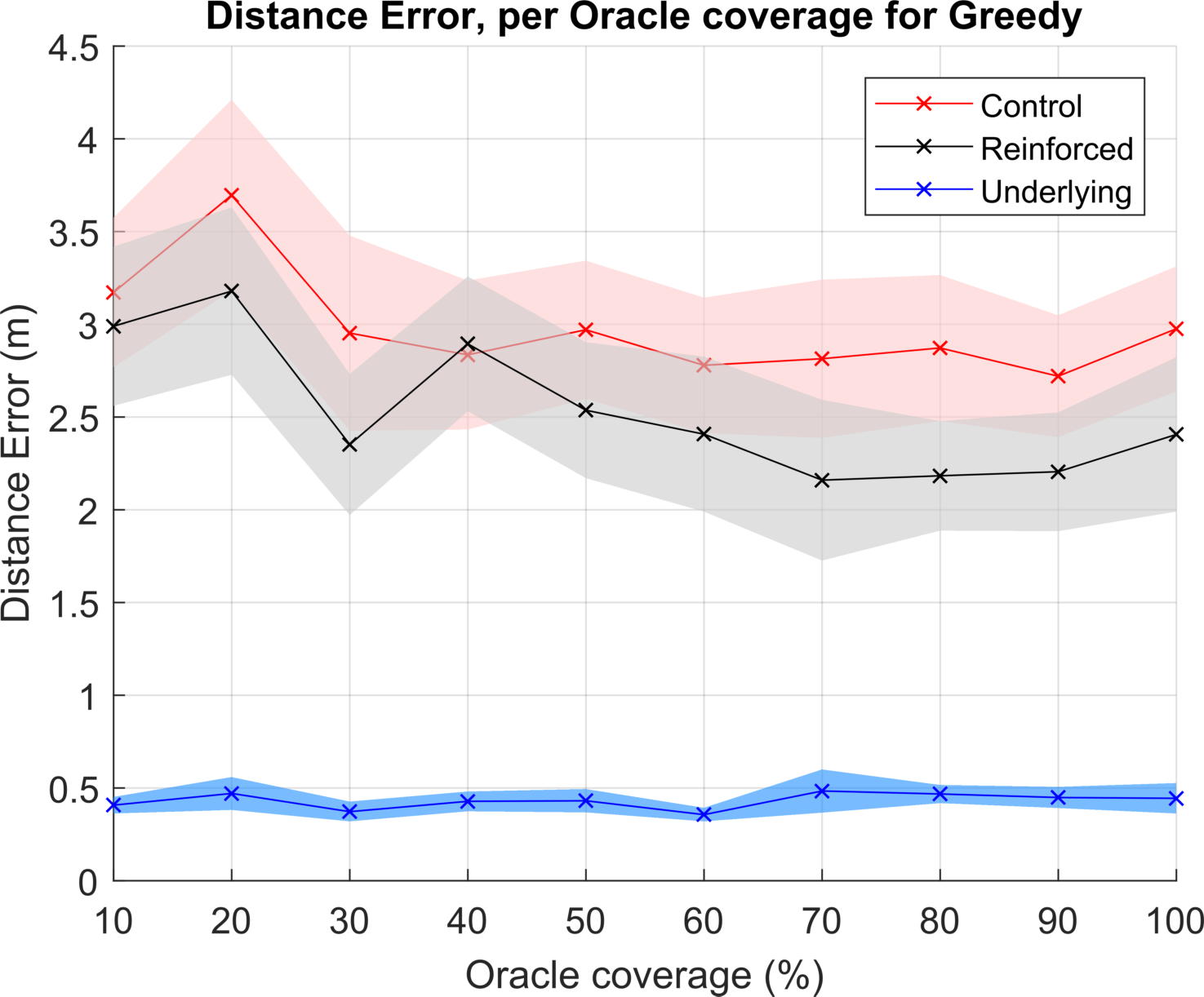}
  \caption{Distance error per oracle coverage under Greedy regime.}
  \vspace{0.5cm}
  \label{fig:sim_g_ora}
  \centering
  \includegraphics[width=0.6\textwidth]{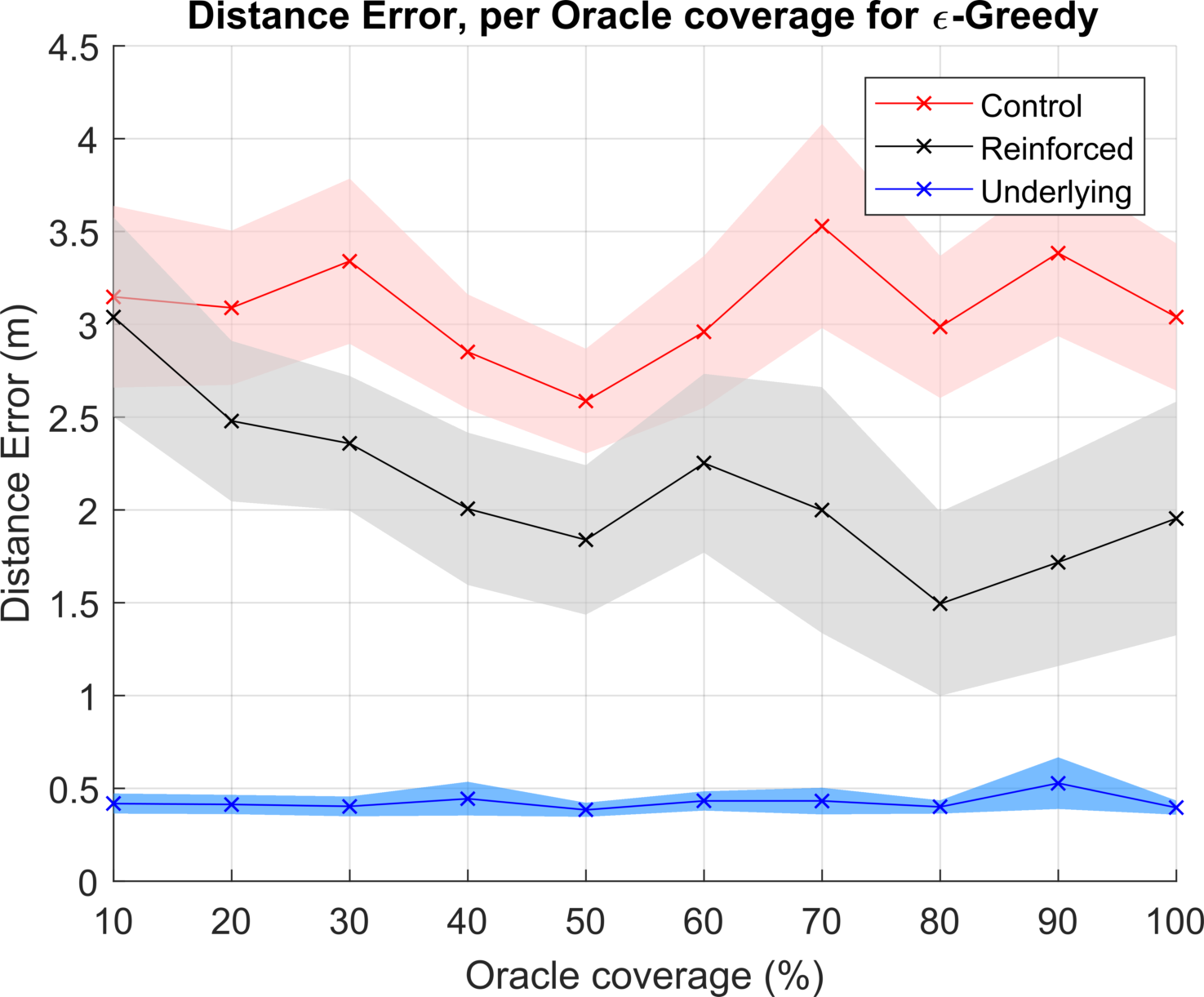}
  \caption{Distance error per oracle coverage under $\epsilon$-Greedy regime.}
  \vspace{0.5cm}
  \label{fig:sim_eg_ora}
  \centering
  \includegraphics[width=0.6\textwidth]{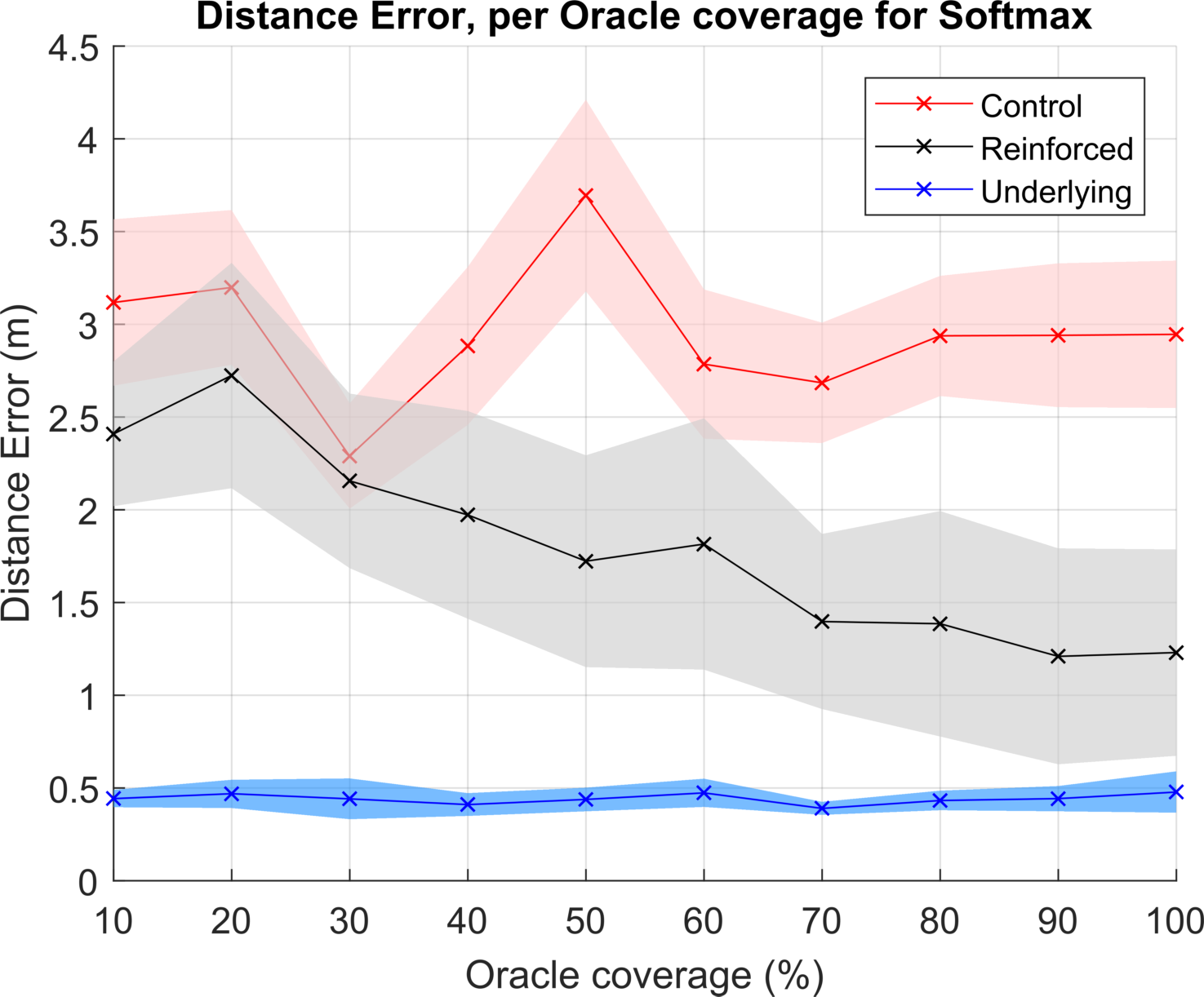}
  \caption{Distance error per oracle coverage under Softmax regime.}
  \label{fig:sim_sm_ora}
\end{figure}

\begin{figure}
\captionsetup{justification=centering}
  \begin{minipage}{0.5\linewidth}
    \centering
    \includegraphics[width=.95\linewidth]{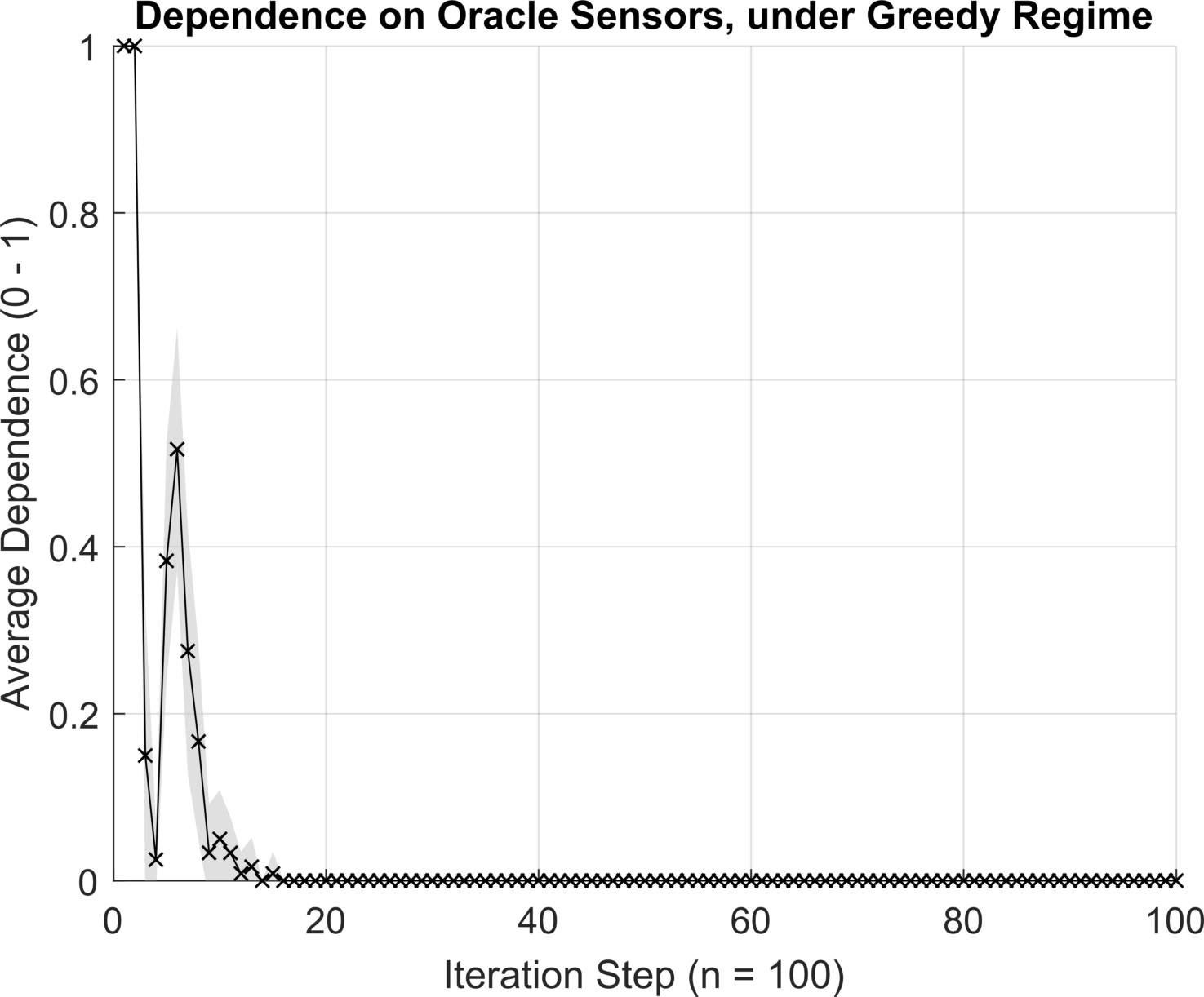} 
    \caption{Oracle Sensors Dependence, Greedy regime} 
    \label{fig:sim_g_dep} 
    \vspace{3ex}
  \end{minipage}
  \begin{minipage}{0.5\linewidth}
    \centering
    \includegraphics[width=.95\linewidth]{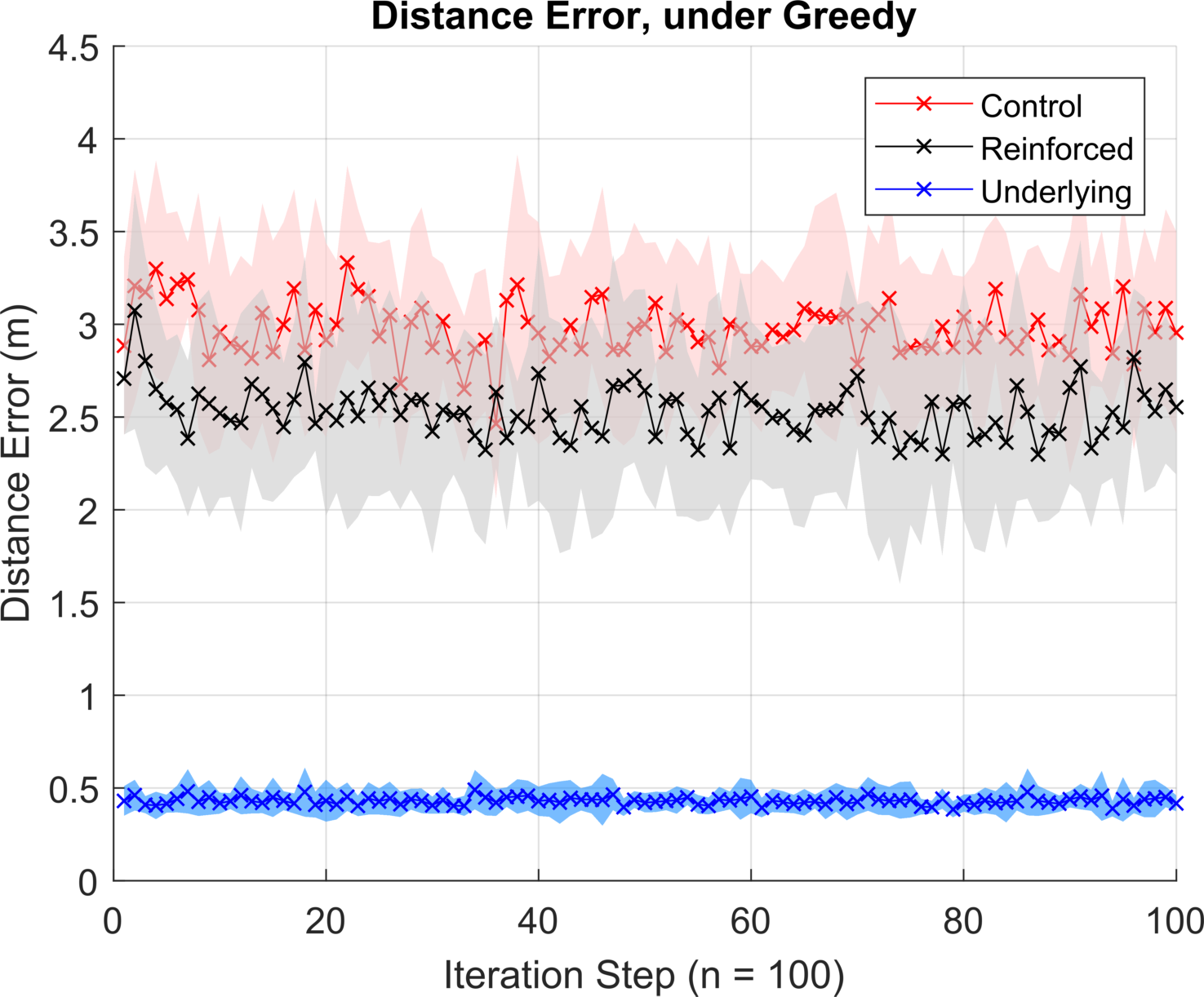}
    \caption{Distance error results, Greedy regime} 
    \label{fig:sim_g_per} 
    \vspace{3ex}
  \end{minipage} 
  \begin{minipage}{0.5\linewidth}
    \centering
    \includegraphics[width=.95\linewidth]{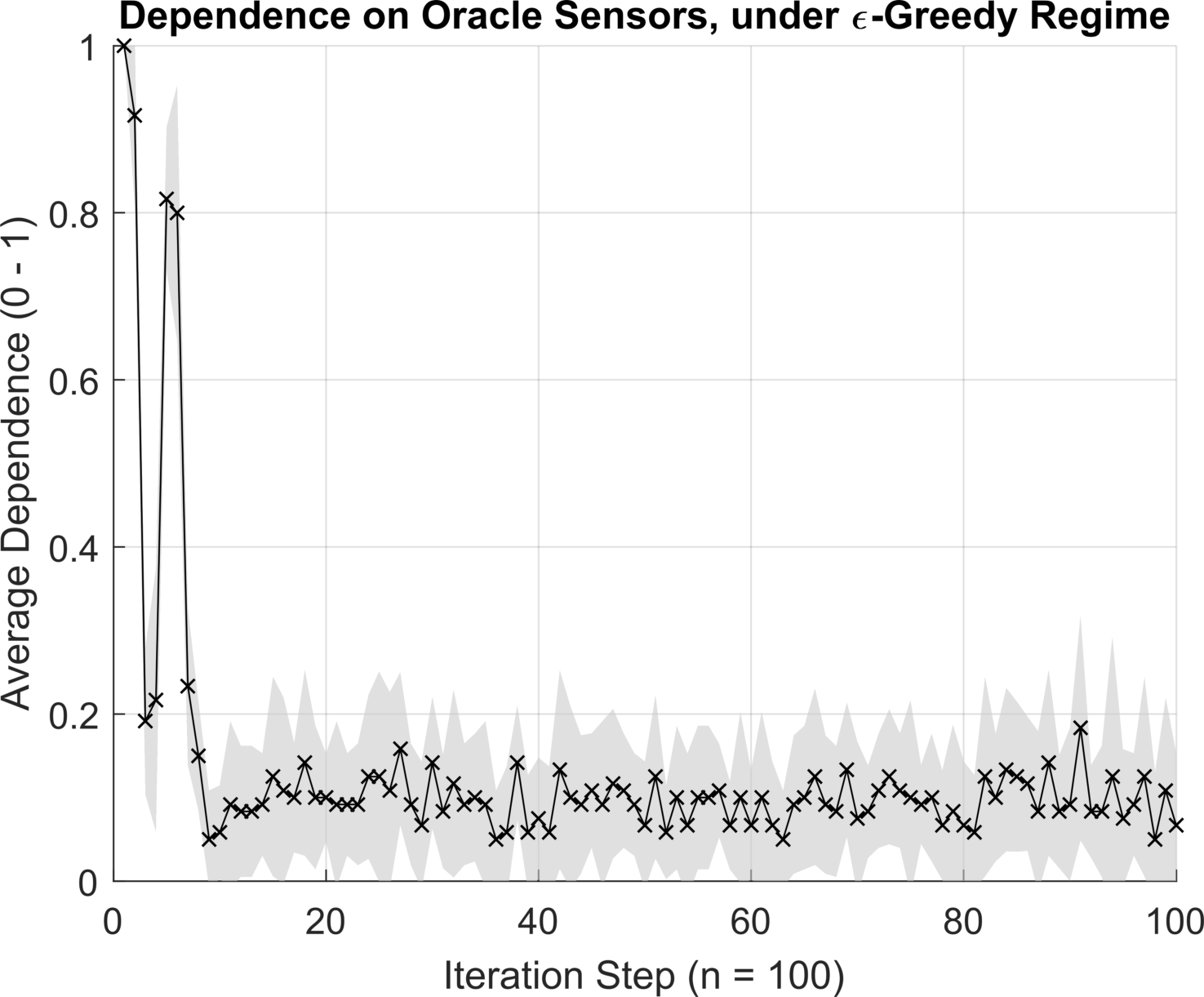}
    \caption{Oracle Sensors Dependence, $\epsilon$-Greedy regime} 
    \label{fig:sim_eg_dep} 
    \vspace{3ex}
  \end{minipage}
  \begin{minipage}{0.5\linewidth}
    \centering
    \includegraphics[width=.95\linewidth]{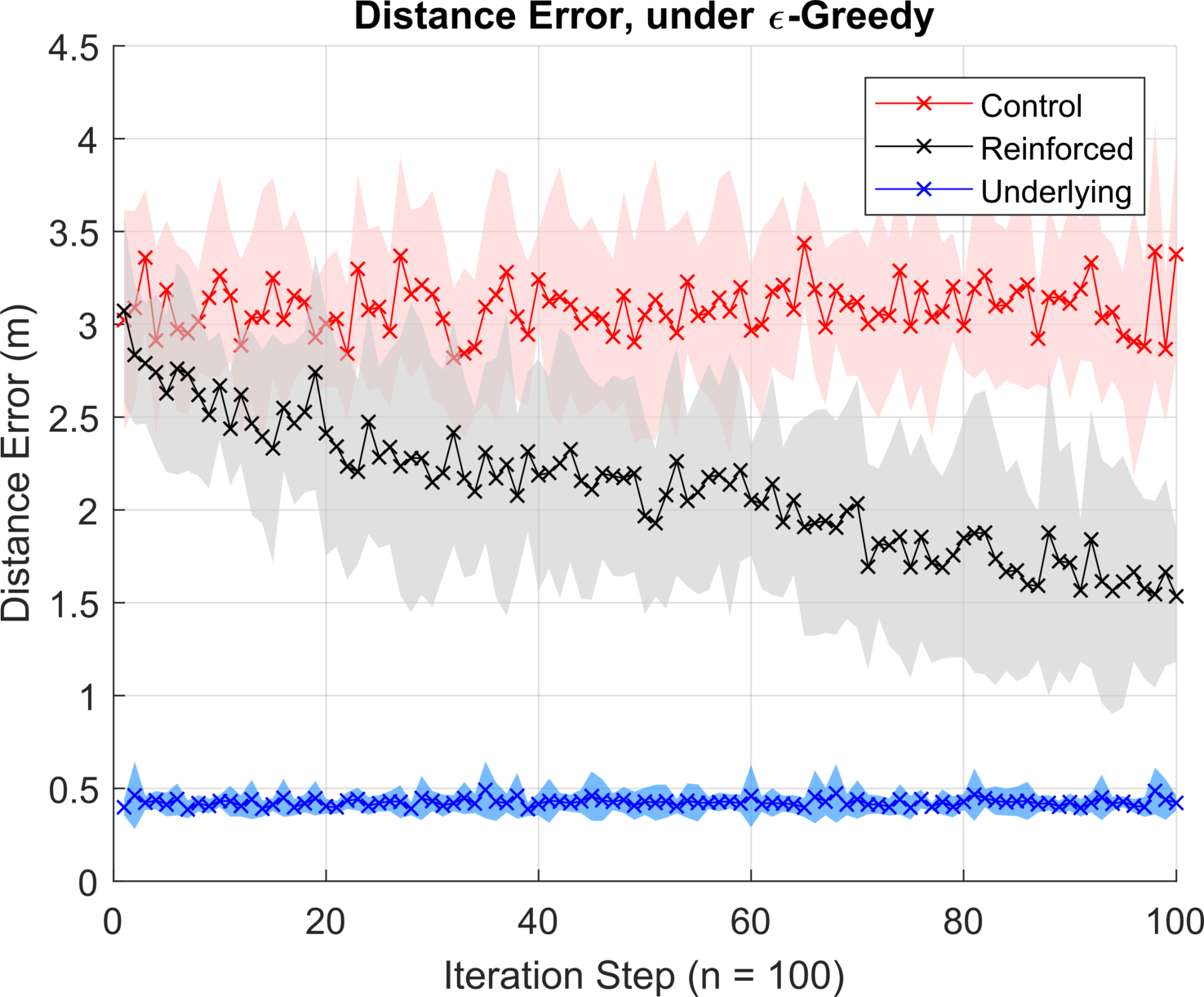}
    \caption{Distance error results, $\epsilon$-Greedy regime}
    \label{fig:sim_eg_per}
    \vspace{3ex}
  \end{minipage} 
  \begin{minipage}{0.5\linewidth}
    \centering
    \includegraphics[width=.95\linewidth]{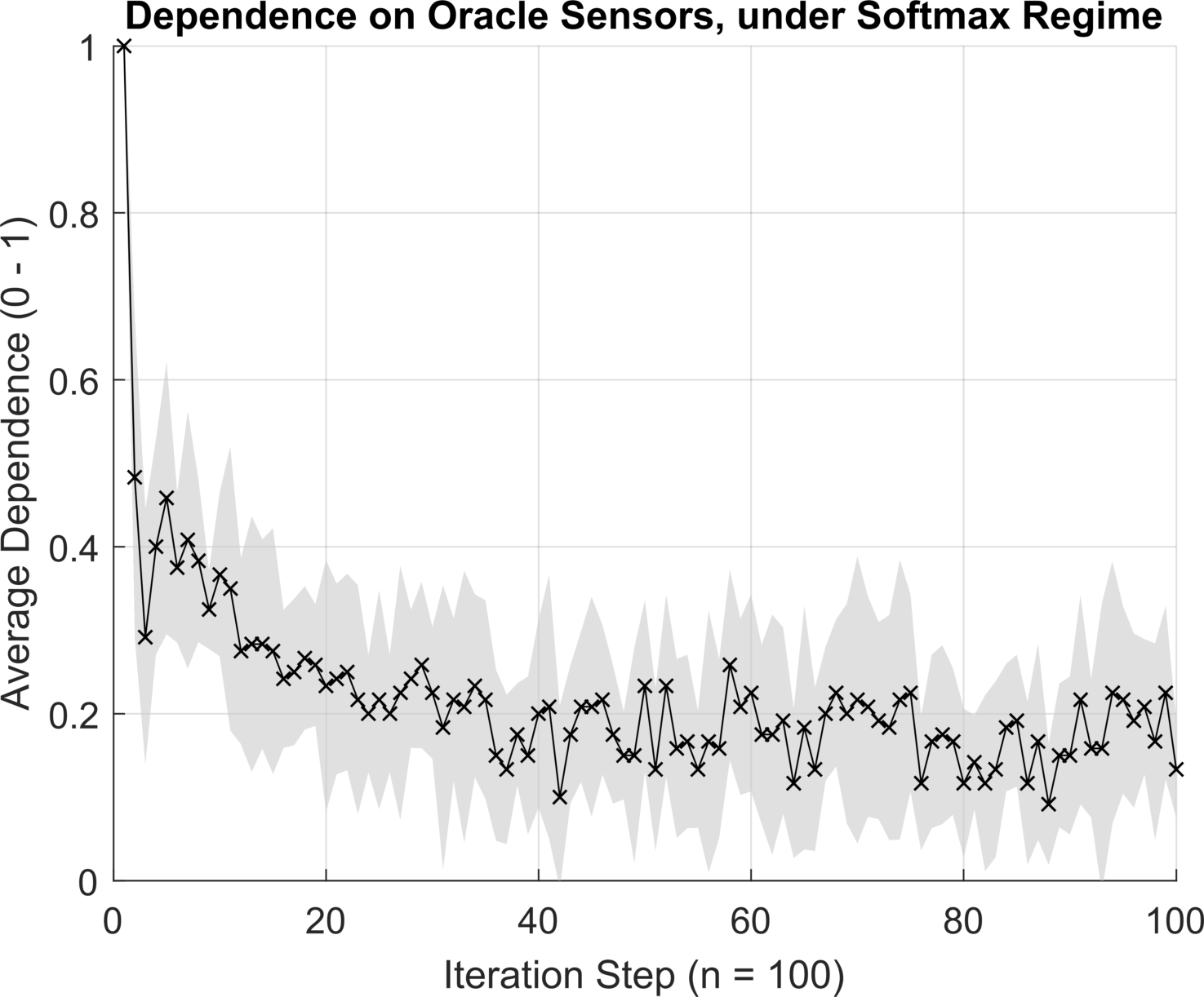}
    \caption{Oracle Sensors Dependence, Softmax regime} 
    \label{fig:sim_sm_dep} 
    \vspace{3ex}
  \end{minipage} 
  \begin{minipage}{0.5\linewidth}
    \centering
    \includegraphics[width=.95\linewidth]{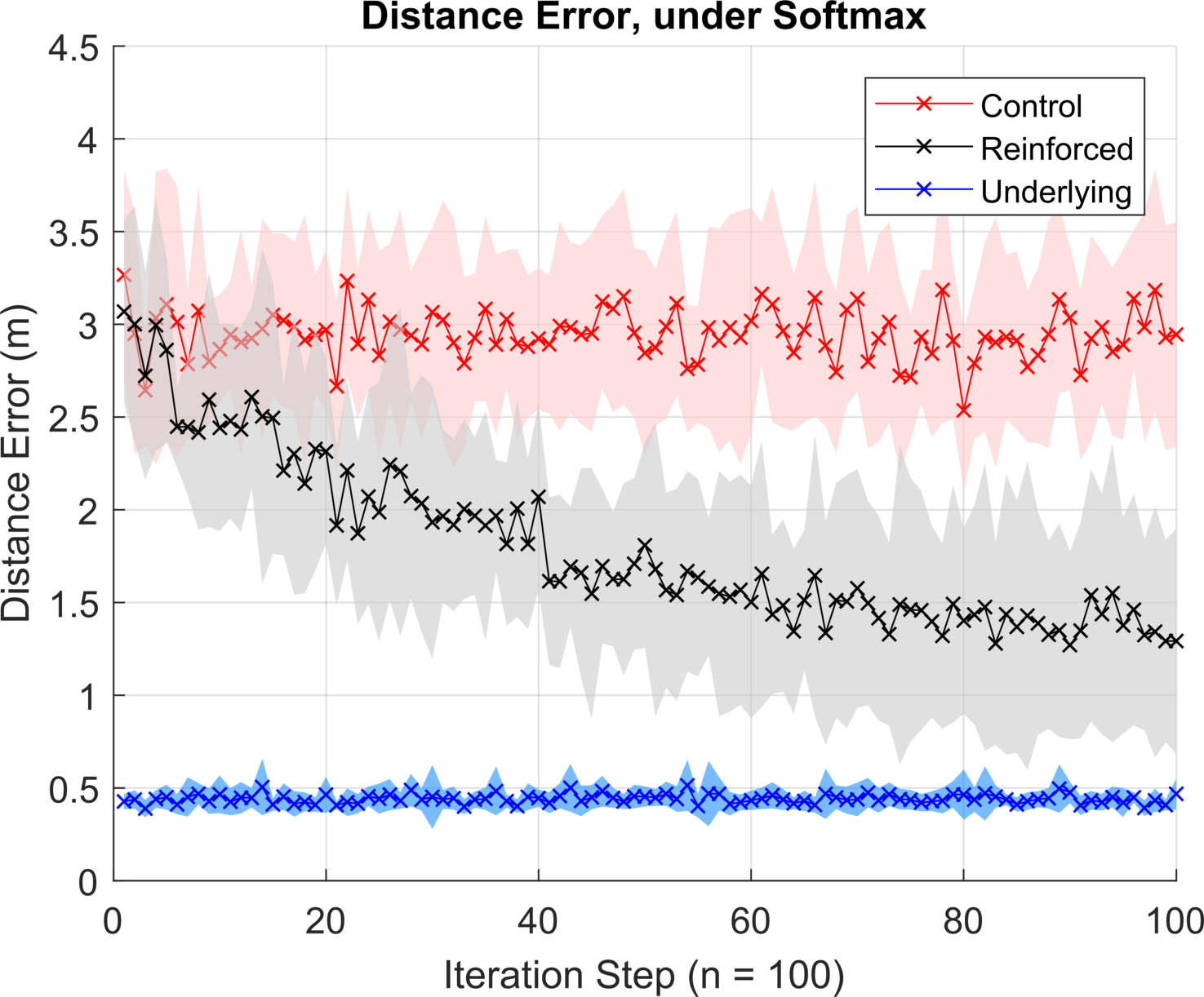}
    \caption{Distance error results, Softmax regime} 
    \label{fig:sim_sm_per} 
    \vspace{3ex}
  \end{minipage} 
\end{figure}

The 3 curves presented in Figs. \ref{fig:sim_g_ora}, \ref{fig:sim_eg_ora}, \ref{fig:sim_sm_ora}, \ref{fig:sim_g_per}, \ref{fig:sim_eg_per}, \ref{fig:sim_sm_per}, are dubbed Control, Reinforced and Underlying. The Underlying curve shows the result of the fundamental distributions which were generated when the synthetic state space was created. They describe the underlying model of the simulated state space, and can be thought of as a localisation result under optimal policy $\bm{\pi^*}$. The Control curve show the result of the the same fundamental distributions, albeit with 3dB of Additive White Gaussian Noise (AWGN) added. This simulates a noisy channel in the indoor environment. The Reinforced distribution is regulated by the presented method. The Control and Reinforced models begin as one and the same. The objective to observe is the Reinforced curve tending towards the Underlying curve as the number of plays is increased, effectively showing how close the algorithm is to the optimal model.

The metric used to measure the performance of the algorithms is the distance error. The distance error is defined as the shortest Euclidean distance between prediction and label in Cartesian space. This metric has previously been used in \cite{kozlowski_data_2018} and is considered as standard in the literature \cite{lymberopoulos_microsoft_2017}. The metric used to show the dependence on energy-inefficient sensors is the total number of iterations where the system stayed in $S1$, divided by the total number of iterations. The normalisation of this metric allows us to represent the dependence as a variable between $[0, 1)$. The closer to 1, the more dependent the system is on multiple sensors. The simulation MDP parameters of $\alpha$ and $\gamma$ were set to 0.4 and 0.9 respectively, and the oracle weight during each re-estimation is 0.7. All of the above were chosen empirically, as they were found to provide best localisation results.

Considering the results of `oracle' usage, the graphs in Figs. \ref{fig:sim_g_ora}, \ref{fig:sim_eg_ora}, \ref{fig:sim_sm_ora} show the average performance of the algorithm as a function of `oracle' coverage. These graphs confirm that the algorithm is viable - the performance of the algorithm under different action selection regimes was compatible with the prediction. The improvement in performance is a function of how much state space is covered by `oracles'. Depending upon which action selection method is chosen, the improvement varies between 0.5m for Greedy to 1.5m for Softmax.

We will now discuss the dependency results from the simulated environment. The graphs are consistent with the hypothesised effect of the action selection method. The 3 Figures, \ref{fig:sim_g_dep}, \ref{fig:sim_eg_dep} and \ref{fig:sim_sm_dep} show the results of Greedy, $\epsilon$-Greedy and Softmax selections respectively. Greedy selection exploits the rewards immediately, and converges to near 0 dependence on `oracles'. This in turn shows, that the closer we are to 0, the faster the distance error converges in Fig. \ref{fig:sim_g_per}. The convergence here is sub-optimal, as the trajectory of the system could be improved. This is visible in Fig. \ref{fig:sim_sm_per}. The Softmax method was used with a temperature of $\tau = 1$. This is displayed with a dependence graph in Fig. \ref{fig:sim_sm_dep}. A gradual roll-off improves the localisation performance. However, this results in higher energy usage, as the energy-heavy senor usage converges to 0.2. 

Figures \ref{fig:sim_eg_dep}, \ref{fig:sim_eg_per}, \ref{fig:sim_sm_dep} and \ref{fig:sim_sm_per} show the optimal trade-off between energy efficiency and quickness of training. The $\epsilon$-Greedy algorithm quickly reaches the mean of 0.1, which is consistent with the parameter $\epsilon$, set to the same value. The Softmax regime provides less erratic reduction of the dependence, in turn providing a better localisation performance. 

The above results are consistent with the theoretical hypothesis. Figs. \ref{fig:sim_g_ora}, \ref{fig:sim_eg_ora}, \ref{fig:sim_sm_ora} confirm, that as the number of oracles increases, so does the improvement in performance. The dependency in \ref{fig:sim_g_dep}, \ref{fig:sim_eg_dep} and \ref{fig:sim_sm_dep} agree with the respective action selection methods. The graphs in Figs. \ref{fig:sim_g_per}, \ref{fig:sim_eg_per} and \ref{fig:sim_sm_per} also conform to their respective regimes. All of the methods will be tested on the SPHERE Challenge data for completeness -- however only these two are in contention to see which one is optimal for the use with this algorithm.

\section{Validation \& Results}\label{validation}

\begin{figure}
\captionsetup{justification=centering}
  \begin{minipage}{0.5\linewidth}
    \centering
    \includegraphics[width=.95\linewidth]{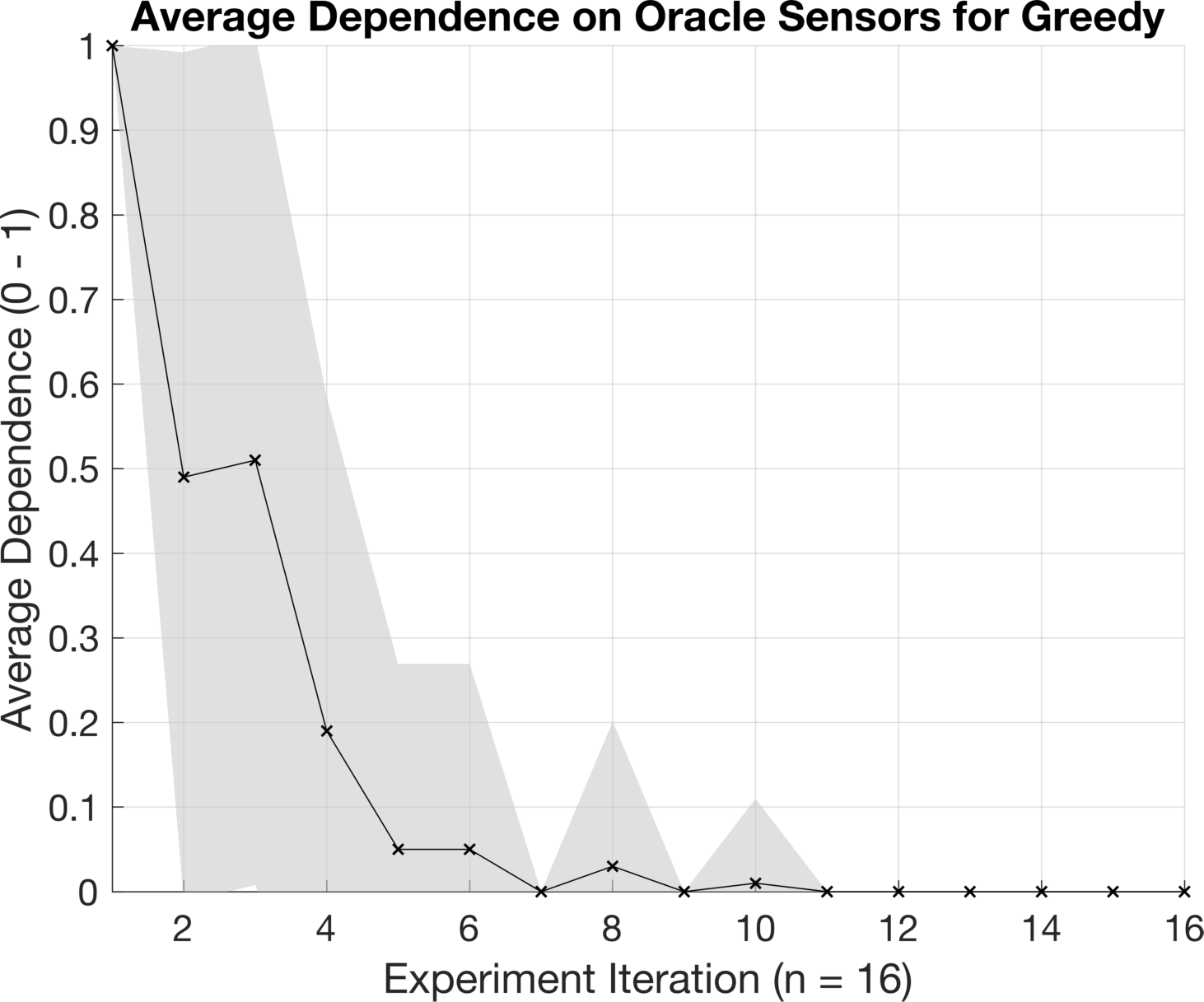} 
    \caption{Oracle Sensors Dependence, Greedy regime} 
    \label{fig:sc_g_dep} 
    \vspace{3ex}
  \end{minipage}
  \begin{minipage}{0.5\linewidth}
    \centering
    \includegraphics[width=.95\linewidth]{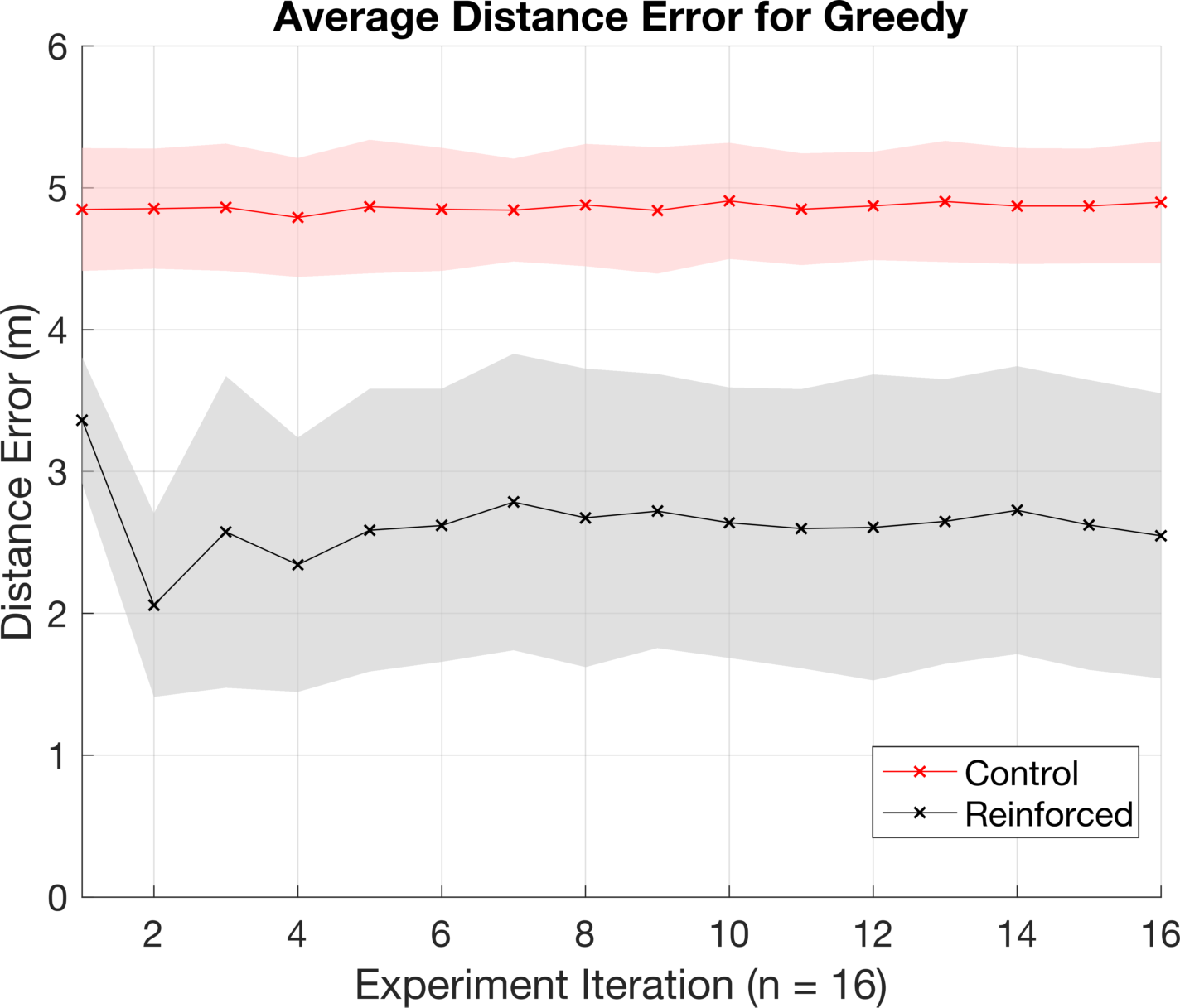}
    \caption{Distance error results, Greedy regime} 
    \label{fig:sc_g_per}
    \vspace{3ex}
  \end{minipage} 
  \begin{minipage}{0.5\linewidth}
    \centering
    \includegraphics[width=.95\linewidth]{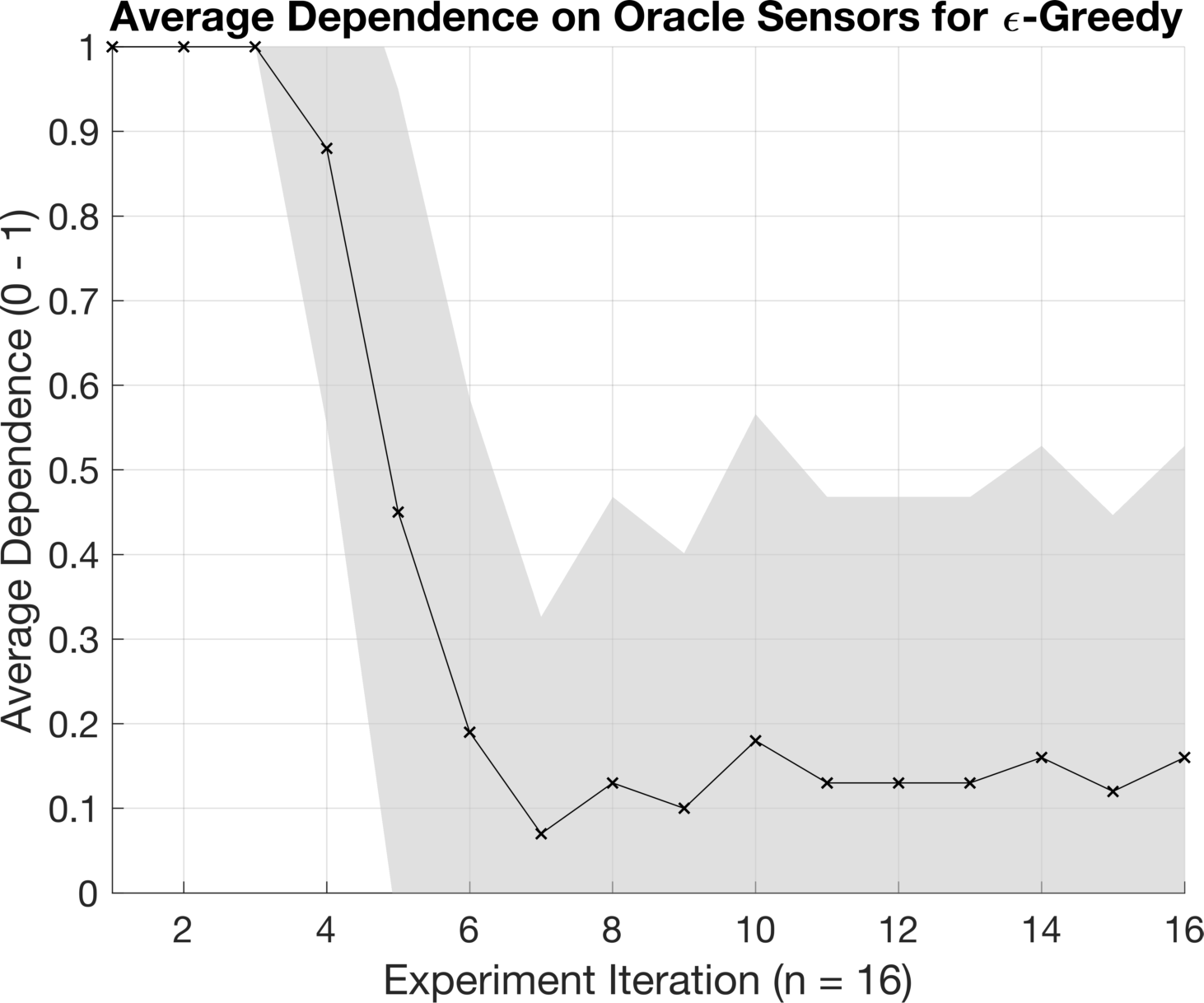}
    \caption{Oracle Sensors Dependence, $\epsilon$-Greedy regime} 
    \label{fig:sc_eg_dep} 
    \vspace{3ex}
  \end{minipage}
  \begin{minipage}{0.5\linewidth}
    \centering
    \includegraphics[width=.95\linewidth]{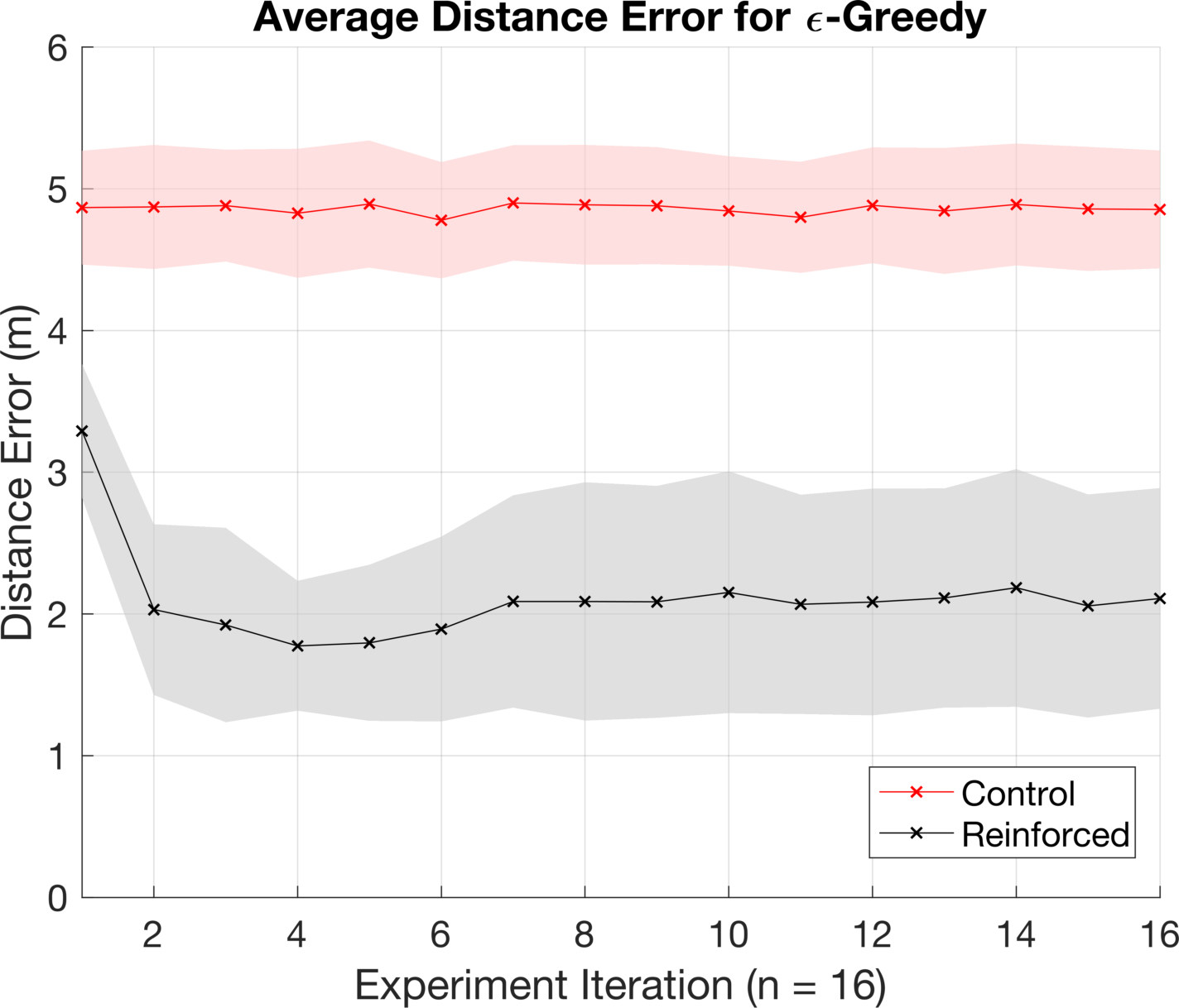}
    \caption{Distance error results, $\epsilon$-Greedy regime} 
    \label{fig:sc_eg_per} 
    \vspace{3ex}
  \end{minipage} 
  \begin{minipage}{0.5\linewidth}
    \centering
    \includegraphics[width=.95\linewidth]{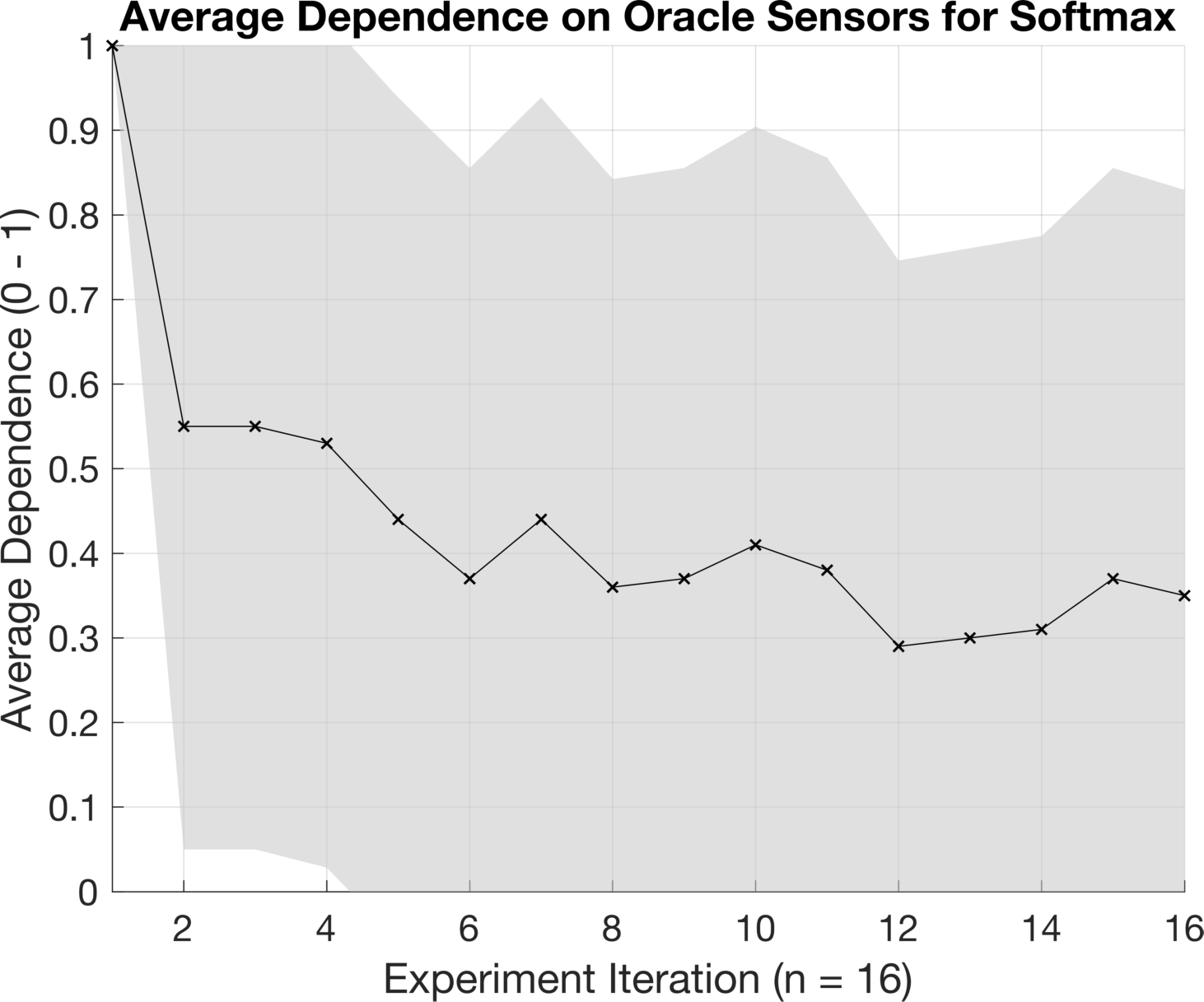}
    \caption{Oracle Sensors Dependence, Softmax regime} 
    \label{fig:sc_sm_dep} 
    \vspace{3ex}
  \end{minipage} 
  \begin{minipage}{0.5\linewidth}
    \centering
    \includegraphics[width=.95\linewidth]{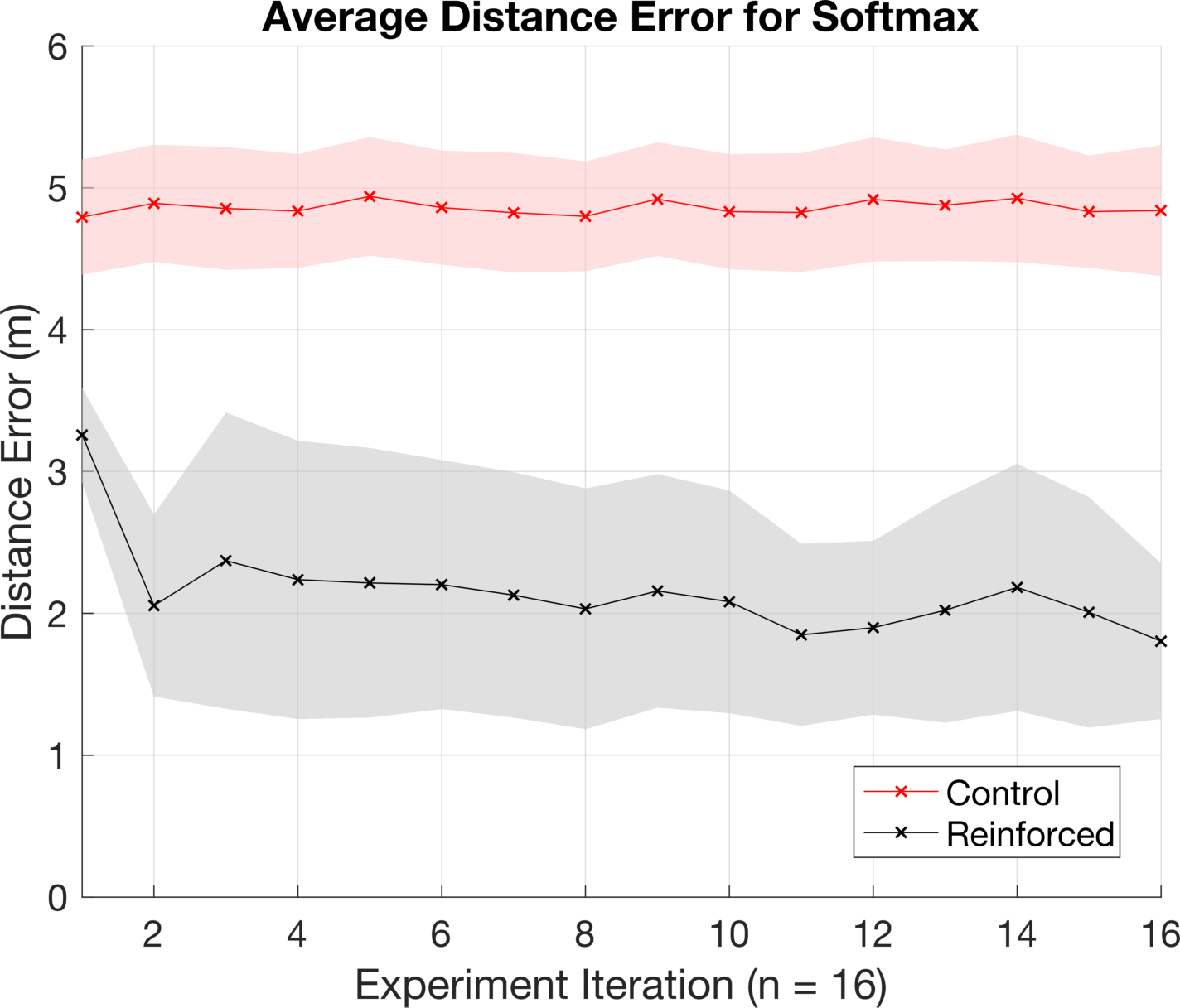}
    \caption{Distance error results, Softmax regime} 
    \label{fig:sc_sm_per}
    \vspace{3ex}
  \end{minipage} 
\end{figure}
The SPHERE Challenge dataset was chosen for this study, as in addition to thoroughly labelled location data, it also includes sensors which can be construed as `oracles'. These sensors include RGB-D video cameras and Passive Infrared (PIR) sensors. The dataset was collected in a 2-storey house currently serving as a pervasive health test bed. The house includes all the amenities and comforts found in a typical residential abode. For further details, we refer the reader to \cite{twomey_sphere_2016}. The version of the dataset used in this study includes labels not available in public domain at the time of publication.

Along with the RSS information available from 4-unique APs scattered around the house, we also have access to room-level location labels. The house includes a total of 9 labelled rooms. No cameras were placed in sensitive locations in the house which meant that some states would lack descriptor `oracles'. The rooms which did include `oracles' are the kitchen, the living room and the downstairs hallway. The PIR sensors however are available in every room. Their usage was limited however, as their reliability was poor. They were used to re-estimate, but were omitted from localisation inference.

The method follows as before. After training a weak model, the energy usage and the relative performance are being scrutinised and leveraged. The data included 19 unique scripted and labelled experiments. In order to obtain a fair result, the training and test proportions were set at 15\% to 85\% respectively. At any one time, a uniform random selection of 3 experiments were chosen to train the model. Testing was performed by running the remainder of user data in a randomly permuted order. This method was repeated $n = 100$ times.

The system was set up such that in $S1$ the system uses a fusion of RSS and camera data. Again, in this state, the parameters are re-estimated according to the labels provided by the `oracles'. As was mentioned above, the PIR were used to re-estimate, but were omitted from location inference. In $S2$, the system relies only on the RSS, with no parameter mixing. The graphs again show the dependency on `oracle' sensors, normalised to 1, and the localisation error convergence graph. The latter further diverges into the control distribution, which is the model trained on initial users and reinforced distribution, which is being continually re-estimated. The localisation labels are room-level.

\begin{table}[!b]
\centering
\caption{SPHERE Challenge performance results}
\begin{tabular}{ c|c|c|c|c }
 \normalsize\textbf{Selection}&\normalsize\textbf{Model}&\normalsize\textbf{25\% of Exp.(m)}&\normalsize\textbf{50\% of Exp.(m)}&\normalsize\textbf{100\% of Exp.(m)}  \\
 \midrule
 \normalsize Greedy             & Control &\normalsize 4.79 ($\pm$ 0.41) & \normalsize 4.87 ($\pm$ 0.43) & \normalsize 4.89 ($\pm$ 0.43) \\
 \normalsize                    & Reinforced &\normalsize 2.34 ($\pm$ 0.90) & \normalsize 2.67 ($\pm$ 1.05) & \normalsize 2.54 ($\pm$ 1.00) \\
 \midrule
 \normalsize $\epsilon$-Greedy  & Control &\normalsize 4.82 ($\pm$ 0.45) & \normalsize 4.89 ($\pm$ 0.42) & \normalsize 4.85 ($\pm$ 0.41) \\
 \normalsize                    & Reinforced &\normalsize 1.77 ($\pm$ 0.45) & \normalsize 2.09 ($\pm$ 0.84) & \normalsize 2.11 ($\pm$ 0.77) \\
 \midrule
 \normalsize Softmax            & Control &\normalsize 4.83 ($\pm$ 0.40) & \normalsize 4.80 ($\pm$ 0.39) & \normalsize 4.84 ($\pm$ 0.46) \\
 \normalsize                    & Reinforced &\normalsize 2.24 ($\pm$ 0.98) & \normalsize 2.03($\pm$ 0.85) & \normalsize 1.80 ($\pm$ 0.55)
 \label{tab:per_results_table}
\end{tabular}
\end{table}

\begin{table}[!h]
\centering
\caption{SPHERE Challenge dependence results}
\begin{tabular}{ c|c|c|c }
 \normalsize\textbf{Selection}&\normalsize\textbf{25\% Iterated}&\normalsize\textbf{50\% Iterated}&\normalsize\textbf{100\% Iterated}  \\
 \midrule
 \normalsize Greedy              &\normalsize 0.19 ($\pm$ 0.39) & \normalsize 0.03 ($\pm$ 0.17) & \normalsize 0 ($\pm$ 0) \\
 \midrule
 \normalsize $\epsilon$-Greedy   &\normalsize 0.88 ($\pm$ 0.33) & \normalsize 0.13 ($\pm$ 0.34) & \normalsize 0.16 ($\pm$ 0.37) \\
 \midrule
 \normalsize Softmax             &\normalsize 0.53 ($\pm$ 0.5) & \normalsize 0.36 ($\pm$ 0.48) & \normalsize 0.35 ($\pm$ 0.48) 
 \label{tab:dep_results_table}
\end{tabular}
\end{table}

The results are presented in Tables \ref{tab:per_results_table} and \ref{tab:dep_results_table}. They show, that the method holds when exposed to non-simulated data. Fig. \ref{fig:sc_eg_dep} shows the dependency graph, with a steady decline in energy-inefficient sensor usage. This can very closely correlate to lower energy consumption. The variability of the dependence was likely caused by the dataset itself. The labels were room-level, which meant that even slight deviation would substantially increase the error. The performance in Fig. \ref{fig:sc_eg_per} also shows a steady performance increase from the control distribution, as the number of plays is being increased.

The data was also scrutinised under the other two action selection regimes. The results for Greedy and Softmax methods are shown in Figs. \ref{fig:sc_g_per} and \ref{fig:sc_sm_per}. The performance of these methods is consistent with the simulated results. For Greedy in Figs. \ref{fig:sc_g_dep} and \ref{fig:sc_g_per}, the system reaches the maximum reward and remains in the `low-power' state. It is also for this method that the performance improvement, relative to the Control model, is the smallest.  

Softmax in Figs. \ref{fig:sc_sm_dep} and \ref{fig:sc_sm_per}, shows a volatile convergence of the dependence. This however translates to a persistent improvement of localisation result. The dependence might be due to the fact that the data used was not as consistent as in the simulation. As it was noted in Section \ref{related}, training and calibration differs for each user. Since this difference was not encoded in the simulation, it could explain the reason for the behaviour of the Softmax method. 

Due to this volatility, as well as higher relative sensor usage in Fig. \ref{fig:sc_sm_dep}, compared with \ref{fig:sc_eg_dep}, the $\epsilon$-Greedy method is the superior method in terms of action selection. This selection method offers a good trade-off between the average sensor usage and the improvement of performance. Whilst the parameter $\epsilon$ was set to 0.1 again, it can be changed depending on the need, allowing it to be more controllable than Greedy or Softmax.

Whilst $\epsilon$-Greedy is the best method for this particular use, the other methods could be advantageous in the context of other sensor applications. Softmax could perform well when considering the amount of sensors used, as opposed to their type, in terms of indoor localisation. In this case, more exploration could translate to better performance, with less regard paid to energy efficiency \cite{li_sensor_2012}. On the other hand, applications where efficiency is critical, could benefit most from Greedy selection. These applications could include on-board feature extraction and activity recognition \cite{elsts2018,esann2018} which would ideally run for a prolonged period of time without recharging.

\newpage
\section{Conclusion}\label{conclusion}

This study has shown a novel adaptive technique for energy-aware indoor localisation. A simulated environment was built and scrutinised against different methods of action selection. A widely available dataset was then used to validate the hypothesised performance under data collected in a real pervasive health test bed. This paper shows that the algorithm can generalise well to non-simulated settings and environments, even when using a dataset which was not inherently collected with localisation in mind.

Results show that using as many sensors as possible is not always advantageous. Comparable performance can be achieved with fewer, by employing the presented algorithm. Through the reduction of the dependence on inefficient sensors we can control the environmental impact of the infrastructure. The study also showed the importance of action selection on the effectiveness of the algorithm. This has a direct effect on the performance-efficiency trade-off. The choice of the regime can make it easier for the user to tune the parameters as they see fit, whether they put more emphasis on performance or efficiency. 

Further work will include generalising this method to a different dataset, not necessarily concentrating on the localisation performance. Activity recognition or other data fusion techniques, scrutinised under energy efficiency can also be studied. A better simulation model for these particular problems could also be developed - one which includes dynamic AP selections and `oracles' which are to some degree fallible. Further work could also be performed on how the state space of this algorithm can scale up or down, in terms of the number of states.

\bibliographystyle{splncs04}
\bibliography{ECML2018.bib}

\end{document}